\documentclass[aps,preprint,showpacs,superscriptaddress,groupedaddress]{revtex4}

\usepackage[T1]{fontenc}
\usepackage[latin1]{inputenc}% Iso 88 59-1

\usepackage{verbatim}
\usepackage{amsmath, multirow, amssymb, wasysym, gensymb}
\usepackage{dcolumn}   % needed for some tables
\usepackage{bm}        % for math
\usepackage{amssymb}   % for math
\usepackage{amsmath}

\usepackage{graphicx}
\usepackage{amsfonts}
\usepackage{graphicx}
\usepackage{graphics}
\usepackage[ngerman, english]{babel}
\usepackage{float}
\usepackage{subfigure}
\usepackage{mathrsfs}

\usepackage{color}
\definecolor{carnelian}{rgb}{0.7, 0.11, 0.11}

\begin{document}
\title{Can the Many-Worlds-Interpretation be probed in Psychology?}
\author{Heinrich P\"as}
%\email{}
\affiliation{Fakult\"at f\"ur Physik,
Technische Universit\"at Dortmund, 44221 Dortmund,
Germany}

\begin{abstract}
\noindent 
A minimal approach to the measurement problem and the 
quantum-to-classical transition assumes a universally valid quantum formalism,
i.e. unitary time evolution governed by a  Schr\"odinger-type 
equation. As had been pointed out long ago, in this view the measurement
process can be described by decoherence which results in a "Many-Worlds" or
"Many-Minds" scenario according to Everett and Zeh. A silent assumption for
decoherence to proceed is however, that there exists incomplete information
about the environment our object system gets entangled with in the measurement 
process.
This paper addresses the question where this information is traced out and 
- by adopting recent approaches to model consciousness in neuroscience -
argues that a rigorous interpretation results in a modern perspective on
the von-Neumann-Wigner interpretation -- namely that the information that is or is not
available in the 
consciousness of the observer is crucial for the definition of the environment
(i.e. the unknown degrees of freedom in the remainder of the Universe). 
As such the Many-Worlds-Interpretation
while being difficult or impossible to probe in physics may become testable in psychology. 
\end{abstract}
\maketitle

The problem of how to understand the quantum-to-classical transition dates back to the early years of quantum mechanics.
Among the various interpretations of the quantum measurement process the ''Many Worlds" or "Many Minds" interpretations (MWI)
\cite{Everett:1957hd,Zeh:1970zz}
are recently becoming increasingly 
popular
\cite{Tegmark:1997me}. This scenario is attractive as it does not introduce any new elements into the formalism of quantum mechanics which go
beyond the 
unitary evolution governed by a Schr\"odinger-type equation, and can in this sense be understood as ''minimal'' or ''conservative''.

In the following we briefly review the measurement process and the role of decoherence in the MWI, before we concentrate on the
crucial importance of the observer's perspective.

In quantum mechanics, the state vector $|\Psi \rangle $ describes the complete knowledge about a system 
and evolves deterministically according to a Schr\"odinger-type equation
\begin{equation}{}
i \frac{d}{dt} |\Psi (t) \rangle = H  |\Psi (t) \rangle.
\end{equation}
The state vector 
does not, however, unambiguously determine the outcomes of measurements. What happens during a measurement can be easily illustrated
by adopting a spin-1/2 or Q-bit object system $| \Psi \rangle \sim \{ |\uparrow \rangle , | \downarrow \rangle \}$ which is coupled to a measurement apparatus with
states $\{ |0 \rangle, | + \rangle, | - \rangle \}$. In the measurement process the object system gets entangled with the apparatus and unitary evolution
yields a state vector $| \Psi_{\rm tot} \rangle$ for the complete (object + apparatus) system
\begin{eqnarray}{}
| \Psi_{\rm tot} \rangle \sim | \uparrow 0\rangle \rightarrow  | \uparrow +\rangle  \nonumber \\
| \Psi_{\rm tot} \rangle \sim | \downarrow 0\rangle \rightarrow  | \downarrow -\rangle. 
\end{eqnarray}
In general our object system will be in a superposition $ |\Psi \rangle \sim \alpha_{\uparrow} | \uparrow \rangle 
+ \alpha_{\downarrow} | \downarrow \rangle$ which in the measurement process then evolves via
\begin{equation}
| \Psi \rangle \sim \alpha_{\uparrow} | \uparrow, 0 \rangle 
+ \alpha_{\downarrow} | \downarrow, 0 \rangle \rightarrow 
| \Psi \rangle \sim \alpha_{\uparrow} | \uparrow,+ \rangle 
+ \alpha_{\downarrow} | \downarrow, - \rangle.
\end{equation}
The latter two terms then correspond to the infamous Everett branches, ''Many Worlds'' or ''Many Minds'': if the apparatus is understood as observer the experimentalist will find herself within one of the branches and will not experience any alternative realities. The understanding of the measurement process as a coupling of object system, measurement apparatus and observer is known as the von-Neumann chain
\cite{Neumann}, and it leads immediately to a new problem: where in this chain is the
quantum-to-classical transition, also known as ''Heisenberg cut'', happening? In principle there could always be an outside observer
(''Wigner's friend'') who would experience her experimentalist buddy in a quantum superposition or ''Schr\"odinger cat state''. 
All that seemed to be known for sure is that
the mind of the observer is in a well-defined state (''psycho-physical parallelism'' \cite{Neumann}). 
In this situation Wigner advocated a quantum collapse triggered in the experimentalist's consciousness 
(''consciousness as the last observer'') and this understanding is known as the Wigner-von-Neumann interpretation
\cite{Wigner}.

Wigner later on changed his mind, however, when H.D. Zeh discovered decoherence
(\cite{Zeh:1970zz}, see also \cite{Joos:1984uk,Zurek:1991vd,Schlosshauer:2003zy,Schlosshauer-Buch}): 
the phenomenon that entanglement with the environment leads to an extremely fast suppression of interference terms/quantum superpositions. Decoherence can be understood most easily in terms of
reduced density matrices $\rho^{\rm r}$. If we consider the density matrix of a quantum system $\rho =  | \Psi \rangle \langle \Psi |$ in the basis 
$| a \rangle$, the matrix element $\rho_{a'a}$ is given by the product of wave functions
\begin{equation}{}
\rho_{a'a} =  \langle a' | \Psi \rangle \langle \Psi | a \rangle \equiv \psi^*(a') \psi(a)
\end{equation}
and a total system consisting of two entangled sub-systems is described by
\begin{equation}{}
\rho_{a'b'ab}
 = \psi^{*}(a',b')
 \psi (a,b).
\end{equation}
Considering now an observable acting only on the $a$-subsystem $L_{a'b'ab}\equiv L_{a'a} \delta_{b'b}$, one can calculate the expectation value
as 
\begin{equation}{}
\langle L \rangle = Tr (\rho L) = \sum_{a'b'ab} \psi^{*}(a',b') L_{a'a}  \delta_{bb'} \psi(a,b) \equiv \sum_{a'a} \rho^{\rm r}_{a'a} L_{a'a}
\end{equation}
with
\begin{equation}{}
\rho^{\rm r}_{a'a}= \sum_{b'} \psi^*(a',b') \psi(a,b').
\end{equation}
By looking only at the $a$-subsystem we trace out or average over the uninteresting degress of freedom of the remainder of the system. The resulting reduced density matrix $\rho^{\rm r}$ looks then exactly like a mixed state.
   
While this process provides an elegant and minimal explanation of the quantum measurement process, it also has two disturbing consequences:

\begin{itemize}

\item
''Many Minds'': the observer ''splits'' into multiple copies observing each possible outcome.
\item
Classical reality is a consequence of perspective. It results from our ignorance about the exact state of the environment.
It is ''emergent'', not fundamental. 

\end{itemize}   

While heated debates have been fought out over the first of these points, this paper argues that the real ''elephant-in-the-room'' is the second consequence: Classical reality is a consequence not only of a measurement system being coupled to an environment but also of the
incomplete knowledge about this environment, which is of course a consequence of the local observer who simply cannot have all possible information about the exact state of the entire Universe. This local perspective has been dubbed ''frog perspective'' (local, classical)
by Tegmark and Zeh in contrast
to the ''bird perspective'' (non-local, quantum) in which the entire quantum system would be observed and no quantum-to-classical transition takes place: The 
quantum-to-classical transition is perspectival!

Thus in principle there are two possible kinds of quantum systems:

\begin{itemize}
\item
Isolated (typically microscopic) systems with no interaction with the environment.
While all quantum systems we have experience with are of this type, this is naturally always an approximation.

\item
The entire quantum Universe: global, encompassing, with no external environment and thus not subject to decoherence. 
 It is this latter system which constitutes the only true fundamental quantum state
 which can be experienced only in the
non-local bird perspective.

\end{itemize}

As a side remark, let me mention that
the quantum Universe thus has a vanishing von-Neumann entropy. It thus seems likely that 
the quantum Universe by itself is timeless, as described by the Wheeler-DeWitt equation \cite{DeWitt:1967yk}. 
It has been argued by Zeh 
\cite{Zeh:1986ix}
and Kiefer \cite{Kiefer:2009tq} that time itself then could be an emergent property of classical spacetime as a consequence of decoherence
related to averaging out irrelevant gravitational degrees of freedom such as gravitational waves or tiny density fluctuations.
In this case, not only the classical world but even time would be perspectival (compare the recent work by Rovelli 
\cite{Rovelli:2015dha}
which argues along these lines).

The main open question of this approach seems to be what defines the ''frog perspective'' or - in Tegmark's words - the ''factorization'' 
\cite{Tegmark:2014kka} into subject, object and environment. Obviously this perspective is a consequence of the observer's consciousness being confined to her cerebral cortex.
But what defines the boundaries of this ''cognitive self'' without assuming a classical description and preferred basis beforehand?
In the following we argue that the ''frog perspective'' may be a crucial prerequisite of consciousness itself. 

Of course the phenomenon of consciousness is far from being understood. In the recent years however an interesting approach called ''Integrated 
Information Theory'' \cite{IIT} has been developed by neuroscientist Giulio Tononi, which pursues a mathematical framework for evaluating the
quality and quantity of consciousness based on properties of the corresponding information processing such as the irreducibility into subsets or ''integration of information''. While such a property reminds of entangled quantum systems, Max Tegmark has provided two important results
which suggest that consciousness at least in the IIT framework can most probably be no quantum process:

\begin{itemize}

\item
In \cite{Tegmark:1999yf} Tegmark estimated decoherence times of neurons and microtubules within the human brain and found that quantum superpositions
decay on extremely fast time scales of the order of $10^{-13}-10^{-19}$ seconds.

\item 
More recently Tegmark applied IIT to quantum systems and found that due to the free choice of the Hilbert space basis
only an insufficient maximum integrated information of 0.25 bits can be obtained \cite{Tegmark:2014kka}.

\end{itemize}

While these results are not unchallenged, we nevertheless thus adopt as a working hypothesis that consciousness should be understood as a
classical algorithm operating in the cerebral cortex and defining the factorization into subject/conscious self, object and environment.
In fact, taking Tononi and Tegmark seriously, consciousness seems to be emergent itself and only possible within a classical perspective.
Consciousness may actually be a by-product or even the trigger of the quantum-to-classical transition.
It should be stressed here that there is not only one possible definition of ''self''. For example, biologist Francisco Varela has argued that
organisms have to be understood as a ''mesh of virtual selves'' \cite{Varela}, including the cognitive or conscious self, the immune or body self and so on.
As we claim here, it is of crucial importance to understand which self or selves define(s) the local perspective giving rise to
decoherence and classical reality.
Indeed, situations where the various selves do not coincide seem to be particularly interesting in this regard. Apart from 
autoimmune diseases in which Varela was mainly interested, such situations include 
altered states of consciousness  such as
hallucinogen intoxication (for example under the influence of lysergic acid diethylamide (LSD-25)) where probands report at least from a subjective perspective a dissolution of the mental self and subjective time within a non-local experience \cite{Wittmann}. It seems that these impressions result from a distortion of the information filtering usually proceeding within the thalamus as a consequence of the intoxication. These aspects of hallucinogen experience exhibit interesting (anti-)parallels to the adoption of the local ''frog perspective'' triggering the quantum-to-classical transition: Conversely, here a classical, local self and possibly even time itself emerges,
as a consequence of the neglect of information about the environment.
Given these parallels it seems not to be too far-fetched to speculate whether the local algorithm constituting the conscious self gets so strongly coupled with the environment as a consequence of the hallucinogen intoxication that it is lifted to a less local perspective
and in this way is able to experience some kind of ''quantum holism''.
One should seriously scrutinize whether such considerations can easily be disregarded as ''New-Age-Bullshit''.

More concretely,
the question proposed to study here is what happens in mental states where consciousness fades away and is experienced as "dissolving".
As Tegmark has argued in \cite{Tegmark:1999yf}, classical systems exist in some kind of intermediate or critical regime between too little interaction (no decoherence) and too much interaction with the environment, where in the latter case it is not meaningful to consider the system as an independent entity anymore. If the "conscious self" defines the local perspective that leads to the perception of a classical world, it is the local and classical property of this "self" which is crucial for the quantum-to-classical transition.
What is proposed here is   
to study the conditions where consciousness looses its independence by increasing the information flux into the algorithm assumed to be its physical correlate.

By adopting these ideas one could come to the fascinating conclusion that while the interpretation of quantum mechanics in general and the
MWI in particular are notoriously difficult (if not impossible) to test in physics experiments, such tests may be possible in psychology.
A simple setup could for example employ probands under the influence of LSD-25 performing quantum measurements (such as spin-up versus spin-down) on a computer screen,
while an equally prepared  control group deals with an equally looking interface connected to a classical simulation based on a random number generator. 
It is conceivable that the first group experiences quantum superpositions while the control group does not.

Obviously, the details of this idea are extremely vague, which is however to some extent a consequence of our limited knowledge about what the physical correlates of consciousness are and how they work. 
Of course it may very well be the case that the algorithm responsible for consciousness remains local even under influence of hallucinogens and that the non-local experience is only an illusion. Moreover, there are open questions about the experimental setup: Is a local perspective already defined by the measurement device or the data processing? Moreover, even a classical random number generator is a quantum physical object on a fundamental level. Can the difference between this object and e.g. a quantum-spin really be exploited this way?
Finally, since even Time itself may cease to exist in this change of perspective, this process is barely understood even from a pure physics perspective. 
These questions cannot easily be answered by theoretical considerations. 
They may be resolved by the proposed experiment, though, in case of a positive result.   

Such a result could be understood as a strong evidence for the MWI and would dramatically affect the way we think about quantum mechanics and
the quantum-to-classical transition, and even about reality itself.

\section*{Acknowledgements}
This work was inspired by discussions with H.D. Zeh, Claus Kiefer, Mathias Becker, Sinan Zeissner and Marc Wittmann as well as recent works by Max Tegmark. 
\cite{Tegmark:1999yf,Tegmark:2014kka}.

\end{document}